\documentclass[aps,amsmath,preprint,amssymb]{revtex4}
\usepackage{amstext}
\usepackage{amsmath}
\usepackage{amssymb}
\usepackage[pdftex]{graphicx}
\usepackage{subfigure}
\usepackage[T1,OT1]{fontenc} 
\usepackage[latin1]{inputenc} \usepackage{amsmath} \usepackage{amssymb}
\usepackage{amscd} \usepackage{graphicx} \usepackage{subfigure}
\usepackage{epstopdf} \usepackage{graphicx}% Include figure files
\usepackage{color}

\begin{document}
\title{Packing of soft asymmetric dumbbells}
\author{An{\fontencoding{T1}\selectfont\dj}ela \v{S}ari\'c\footnote{These authors contributed equally to this work}, Behnaz Bozorgui$^*$, and Angelo Cacciuto}
\email{ac2822@columbia.edu}
\affiliation{Department of Chemistry, Columbia University, 3000 Broadway, \\New York, New York 10027}
%Department of Chemistry, Columbia University, 3000 Broadway, \\New York, New York 10027

\begin{abstract}
We use numerical simulations to study the phase behavior     
of a system of purely repulsive soft dumbbells as a function
of size ratio of the two components and their relative  degree of deformability.
We find a plethora of different phases which includes  most of the mesophases
observed in self-assembly of  block copolymers, but also
crystalline structures formed by asymmetric, hard binary mixtures.
Our results detail the phenomenological behavior
of these systems when softness is introduced in terms of two different classes of
inter-particle interactions: (a) the elastic Hertz potential,
that has a finite energy cost for complete overlap of
any two components, and (b) a generic  power-law
repulsion with tunable exponent.
We discuss how simple geometric arguments can be used to account for the 
large structural variety observed in these systems, and detail the
similarities and differences in the phase behavior for the two classes
of potentials under consideration.
  
\end{abstract}
\maketitle
\section{Introduction}

The problem of optimal packing of spheres in two and three dimensions is a subject that 
has inspired multiple generations of scientists. The first conjectures for equally sized 
spheres were  put forward by Kepler  at the beginning of
the seventeenth century, and were finally proven in 1998 by Thomas Hales~\cite{Hales}.    
Far from being a merely academic problem, packing and self-assembly of nanoparticles and mesoscopic components have
become  major challenges in nanotechnology.
Early experiments on colloidal dispersions~\cite{pusey} have shown how hard spherical particles 
can spontaneously organize into ordered structures when compressed at sufficiently large volume fractions, 
thus providing a microscopic 
realization of the spherical packing problem. The nature of this fluid to solid transition is understood
in terms of the configurational entropy of the system which, at hight densities, is maximized when  
the particles reside on the lattice sites of an FCC or an HCP crystal. 
The relative free energy of  FCC and HCP structures for hard spheres 
was studied in detail by computer simulations~\cite{wilding}. 

Although the specific case of hard spherical particles is a very relevant one, 
as it represents probably the simplest nontrivial mathematical reference model,  
the number of ways in which such particles aggregate into ordered structures is very limited; 
even in the presence of additional isotropic attractions due for
instance to dispersion forces as in the Lennard-Jones interaction or to depletion forces
as in colloid-polymer mixtures.

Recent advances in particle synthesis~\cite{DeVries,Schnablegger,Hong,Weller,Hobbie,weitz,pine,mitragotri} 
have significantly extended the structural landscape via the development of colloidal particles 
that are anisotropic both in shape and surface chemistry, thus
providing an unlimited number of building blocks that can spontaneously organize into an unprecedented variety 
of structures with potentially novel functional, mechanical, and optical properties.
For this reason, the problem of self-assembly of nanoparticles is today under intense investigation. 
Although most of the work has focused on how specific  interactions (typically attractive) may be tailored to  
drive self-assembly of nanocomponents, over the past few years it has been shown how a large number of structures 
with substantially different packing symmetries than FCC, HCP or BCC  may  become accessible  even to
purely repulsive spherical particles, either by altering the shape of their mutual interactions or
 by relaxing the constraint of mutual impenetrability. 
The formation of mesophases by particles interacting via a hardcore plus a repulsive shoulder potential
was analyzed in~\cite{glaser}, and the relative stability of A15, BCC and FCC crystal structures  
as a function of particle softness was discussed in~\cite{kamien}.  
More complex mesoparticles such as  charged or neutral star polymers, 
dendrimers or microgels are usually dealt with by developing a coarse-grained representation 
of the components which are reduced to objects with a simple shape
interacting via ad-hoc effective pair potentials. The study of the physical properties of systems adopting these 
exotic intercations, some of which allowing for even complete overlap among the components, 
has been the subject of several publications~\cite{louis,likos0,likosA,likosB,denton,gottawald,pierleoni, Capone, bozorgui}. These studies unveiled the existance of an unprecedented richness in phenomenological behavior
including the possibility for some classes of interactions to form polymorphic cluster phases~\cite{mladek} 
or reentrant melting transitions followed by multiple transitions between  
ordered crystalline phases~\cite{suto,likos2,pep}, and set the stage for the work presented in this paper.
For a recent review on the subject we refer the reader to reference ~\cite{likos_review}.
  
In our recent work~\cite{pep}, we have shown that a large number of 
solid phases becomes accessible upon increasing the density of a  system of spherical particles interacting
via a Hertz potential. Among the many phases we found crystals with cubic, 
trigonal, tetragonal, and hexagonal symmetries. 
In this paper we go one step further and include an extra degree of complexity to the problem: particle geometry.
Specifically, we explore the relative role of softness and geometry in a system of asymmetric, 
repulsive dumbbells. We  present the system phase behavior as a function of external 
pressure and discuss the interplay between geometry and softness as a means of understanding
the formation of the large number of ordered phases found in the present work.  
We have chosen the asymmetric dumbbellar model because  it allows us to 
efficiently and systematically incorporate both geometrical asymmetry and particle softness into 
the single components. In order to keep our results as general as possible
we decided to describe the interparticle effective interactions using two 
common functional forms: (1) the Hertz potential, 
which describes the energy cost associated with elastic deformation of an amount $h$ between two spheres,
and has the form $V(h)\propto Y h^{5/2}$ ( Y is the Young's Modulus) \cite{landau}, and (2)  
a generic inverse power-law $V(r)\propto r^{-n}$  cut off at the particle diameter. 

In the first case we tune the softness by changing the Young's modulus $Y$, whereas in the second case 
this is done by changing the exponent $n$.

\begin{figure}[h!]
\center
\includegraphics[width=50mm]{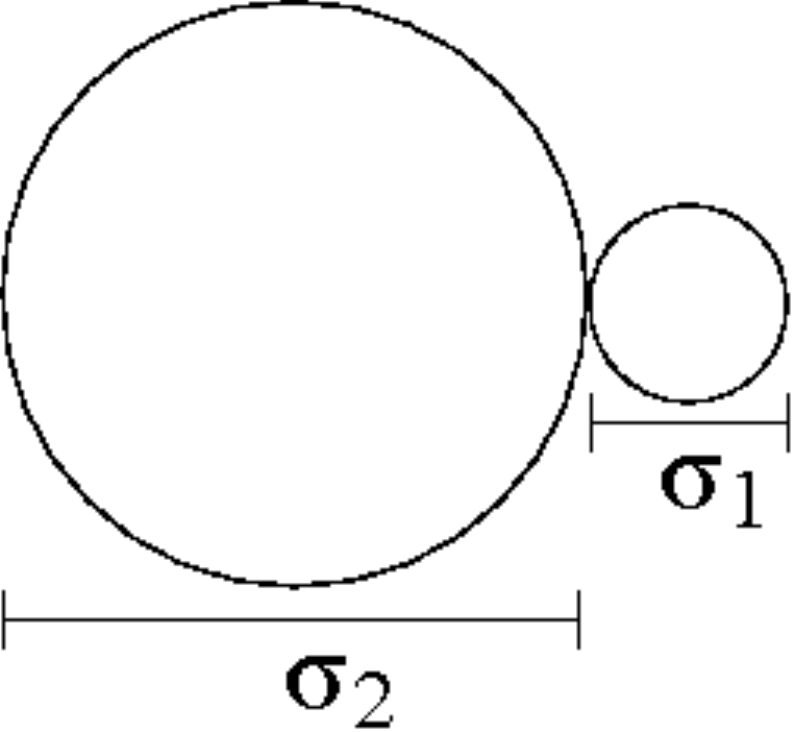}
\caption{Schematic representation of our model for a soft asymmetric dumbbell. Two soft spheres of diameter $\sigma_1$ and $\sigma_2$, with  $s=\sigma_2/\sigma_1 \geq 1$, are connected with a rigid bond of length $d=1/2(\sigma_1 + \sigma_2)$. Both paricles are assumed to have equal mass.} \label{scheme}
\end{figure}

\begin{figure}[h!]
\center
\subfigure[]{\label{fig:edge-a}\includegraphics[width=80mm]{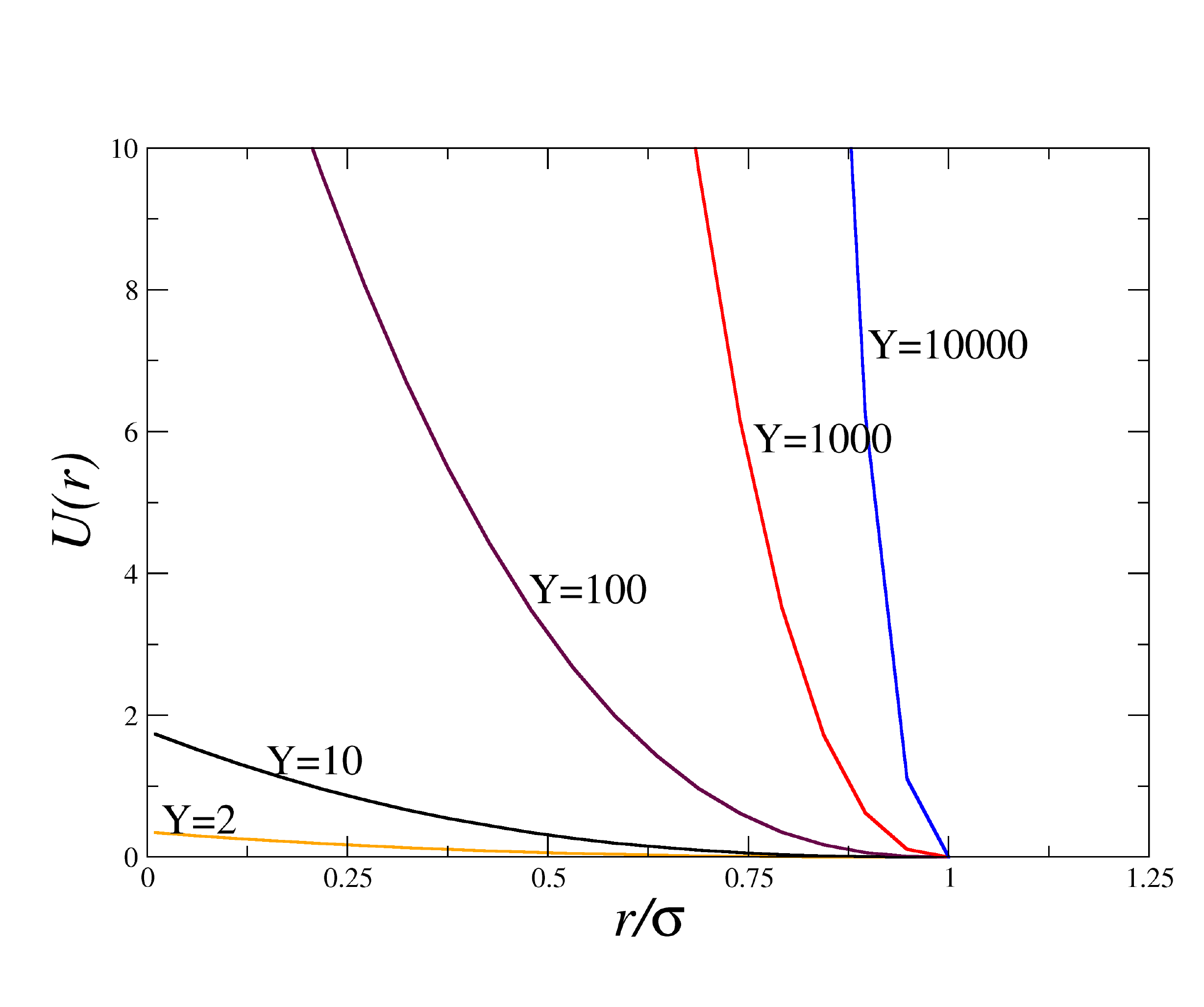}}
\subfigure[]{\label{fig:edge-b}\includegraphics[width=80mm]{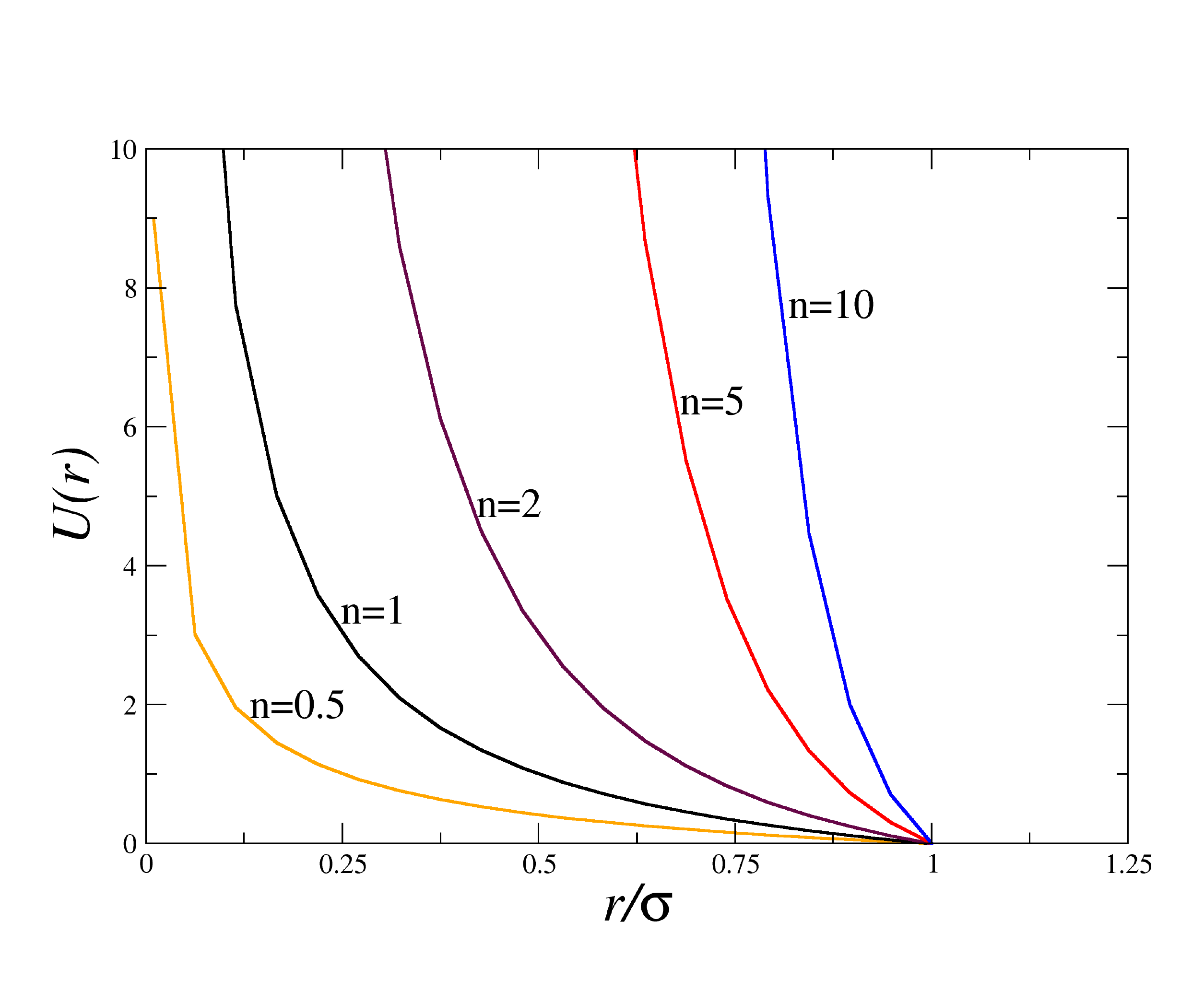}}
\caption{Illustration of the pair potentials used in this work to model the soft interactions between the dumbbellar components. a) The Hertz potential according to Eq. (1), plotted for the range of values of the Young's modulus $2 \leq Y \leq  10^4$. b) The generalized inverse power law potential (GIPL) according to Eq. (2), plotted for the range of the exponents $0.5 \leq n \leq 10$.} \label{potential_scheme}
\end{figure}

When both tuning parameters tend to infinity, the two models become identical and the hard-repulsive
 limit is recovered. It is important to stress that unlike the inverse power law interaction,
the Hertz potential is bound, i.e. there is a finite energy cost to completely overlap two components. In this 
respect the two potentials can be considered as belonging to different classes of interactions, and we therefore 
discuss them independently. The schematic representation of our model for a soft asymmetric dumbbell is shown in Fig. \ref{scheme}, while the pair potentials we use are illustrated in Fig. \ref{potential_scheme}.

We begin by considering the elastic pair potentials. The dumbbells are modeled as two soft spheres of diameter 
$\sigma_1$ and $\sigma_2$, connected with a rigid bond of equilibrium length $d=1/2(\sigma_1 + \sigma_2)$.
The softness of the spheres is modeled via the Hertz potential, 
that describes the change in the elastic energy of two deformable spheres when subjected 
to axial compression \cite{landau, patricio}. The potential has the following form:
\begin{equation} \label{Hertz}
U_{i,j}(r)= 
\begin{cases}
\frac{32}{45}\frac{Y_{i} Y_{j}}{Y_{i}+Y_{j}}  \left (\frac{\sigma_{i} \sigma_{j}}{ 2(\sigma_{i}+\sigma_{j}) } \right)^{1/2}
 [(\sigma_{i}+\sigma_{j})/2-r_{ij}]^{5/2}  & \text{, $r_{ij} < (\sigma_{i}+\sigma_{j})/2$}\cr
0 &\text{, $r_{ij}\geq (\sigma_{i}+\sigma_{j})/2$}\cr
\end{cases}
\end{equation}
The indices $i,j \in {1, 2}$ indicate the identity of the spheres, $r_{ij}$ is the distance between the centers of any two spheres that belong to different dumbbells, and $Y_{i}$ is the Young's modulus of the sphere. In the equation above the Poisson ratio is taken to be 1/2. When $ Y_{i} Y_{j}/(Y_{i}+Y_{j}) \rightarrow \infty$ the interaction between the two spheres approaches the hard-core limit. When $Y_{i} Y_{j}/(Y_{i}+Y_{j}) \rightarrow 0$ the spheres can completely overlap. The important property of the Hertz potential is that it is bounded and remains finite at $r_{ij}=0$. 

We used the {\sc LAMMPS} molecular dynamics package~\cite{lammps} with a Nos\'{e}/Hoover thermostat in the $NPT$ ensemble to study the phase behavior of a  system of 512 dumbbells for a wide range of system densities, for different values of $Y_1$ and $Y_2$  and several values of size ratio $s=\sigma_2/\sigma_1$. 
 The simulation box was set to be a cube with periodic boundaries for most of our simulations, no difference in the results was found when decoupling the box lengths in each of the three Cartesian coordinates. The system initial configurations were prepared by performing $NPT$ simulations in the gas phase at a very low pressure $p$. Once the system is equilibrated, we slowly ramped the pressure to the desired final value $p_f$ starting form $p_0$ in small increments $\Delta p$. Each subsequent simulation performed at the constant pressure $p_i$ started form the thermalized configuration at the pressure $p_{i-1}=p_i-\Delta p$. This slow pressure annealing procedure should allow the system to fully equilibrate at each pressure. For each $\Delta p$, simulations were run for a minimum of $2 \cdot 10^{6}$ steps with the reduced time-step size $\delta t=0.008t_0$ ( where $t_0$ is the 
 unit time expressed in standard MD units).

Our system has four independent parameters: namely $s$,  $Y_{1}$, $Y_{2}$ and system pressure $p$,
 making the phase diagram  very complex and multidimensional. We find that a convenient way of 
presenting the results is by showing a sequence of slices in the $\{Y_{1},Y_{2}\}$ two-dimensional plane, each at constant $s$ and $p$. 
To further reduce the size of the phase space we impose the following restrictions: $Y_1\geq Y_2$ and $\sigma_1\leq\sigma_2$, 
i.e. the smaller component of radius $\sigma_1$ is always the one that is harder to deform. Given our previous work~\cite{bozorgui}, 
we also expect this constraint to result in the most interesting structure formation.

Finally, all pressures  have been rescaled to properly account for the different sizes of the dumbbellar components, $p^*=p (s^3+1) (\pi\sigma_1^3/6)$. 
Since there are no attractive forces in our system, the formation of these phases is mostly due to the minimization of the repulsion energy as a response to the applied external pressure. However, entropic contributions are not completely negligible~\cite{pep}, especially in the case of dumbbell-crystallization at low pressures, as will be described below. 
 
\begin{figure}[h!]
\center
\includegraphics[width=100mm]{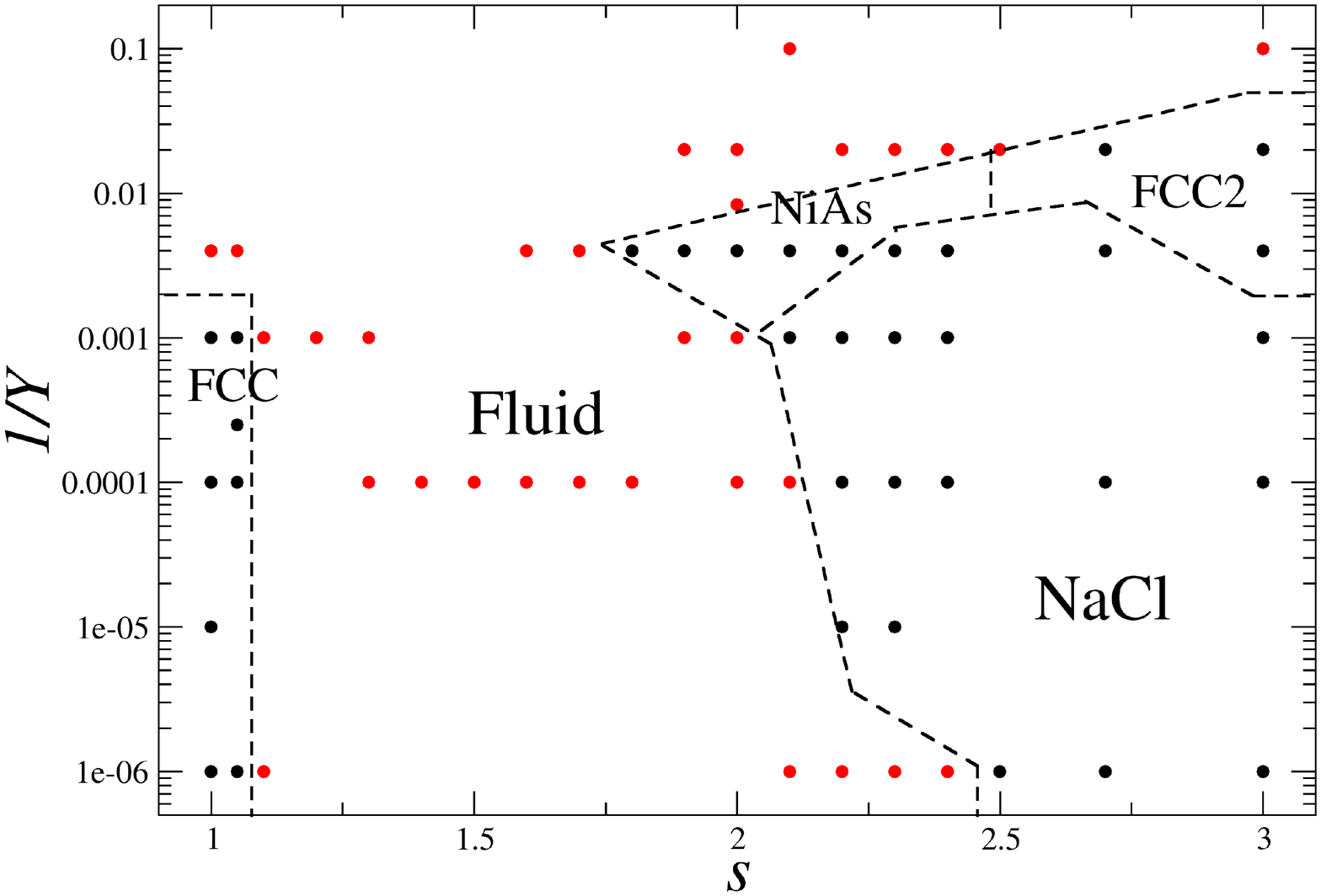}
\caption{Crystallization of Hertzian dumbbells for size-ratios $1 \leq s \leq 3$ and rigidities $Y_1=Y_2\in[10,10^6]$. Below a size-ratio-dependent value of $Y$  
dumbbells stop  crystalizing for any  value of $s$ (ideal gas limit). In the opposite regime (hard dumbbell limit) crystallization 
occurs below $s=1.05$, where the spheres arrange into FCC lattice, and above $s=2.4$, where the dumbbells arrange into an NaCl-type crystal. 
The width of the crystallization gap narrows as the  repulsion become softer and other crystal types become accessible. Here we indicate the
NiAS-type and crystals of the large spheres arranged into an FCC structure (FCC2), which are the ones that first nucleate from the fluid phase. Other structures 
can be formed upon further increase of the pressure (see text for details).} \label{gap}
\end{figure}

Let us start by looking at the hard-limit behavior, i.e. when $Y_1=Y_2 \geq 10^4$, depicted in Fig. \ref{gap}. When the two components have equal diameter (and up to $s=1.05$), the dumbbells, as expected, crystallize into an FCC crystal. We also find that the system can easily crystallize for size ratios in the range $2.4\leq s\leq3$. In this case the dumbbells pack into a NaCl-type binary crystal. Interestingly, we find that for intermediate values of $s$ ($1.05<s\leqslant 2.4$)  the system remains fluid, implying that the hard limit has a crystallization gap in the size ratio space.
These results are a bit surprising as a few papers dealing with the equilibrium properties of binary AB-type hard-sphere mixtures and hard dumbbells~\cite{hudson, dijkstra1, dijkstra2}
report the  stability of a variety of crystalline structures across a comparable range of size ratios. One would expect some of these structures to also nucleate from the fluids 
phase in our system, especially the ones predicted for hard dumbbells~\cite{dijkstra2}. We believe that this is only an apparent discrepancy, and can be resolved in terms 
of the kinetics of the system. In fact, it is well known that the probability of crystal nucleation of hard spheres becomes quickly reduced above a certain degree of polydispersity \cite{frenkel1}.
It is therefore possible that the crystals reported in~\cite{dijkstra2} are indeed the most stable ones. However, it is also possible that for $1.05<s\leqslant 2.4$ they are kinetically inaccessible and for $2.4\leq s\leq3$ 
the closest in free energy to the partent fluid phase is simply the NaCl-type.
What is also very interesting is that as soon as the hard limit is abandoned, the crystal phase space increases in size and diversity: the upper limit of this crystallization gap is narrowed to $s=1.8$, and other crystal structures, besides NaCl-type, are observed, including  NiAs-type binary crystals, as well as the FCC monocrystals of the large spheres. For sufficiently soft particles  ($Y < 50$) the system remains in the fluid state at any pressure.
Notice that Fig. \ref{gap} only reports the identity of the first crystal (the one formed at the lowest pressure) that nucleates from the fluid.
In fact, as the system's Young's modulus is decreased and particles become penetrable a series of different crystalline structures can be generated upon increasing the external pressure.
This peculiar behavior is intrinsic to soft particles~\cite{suto,likos2}, 
and the equilibrium phase diagram  for spheres with this specific potential has been thoroughly described in ref.~\cite{pep}.

\begin{figure}[h!]
\center
\includegraphics[width=180mm]{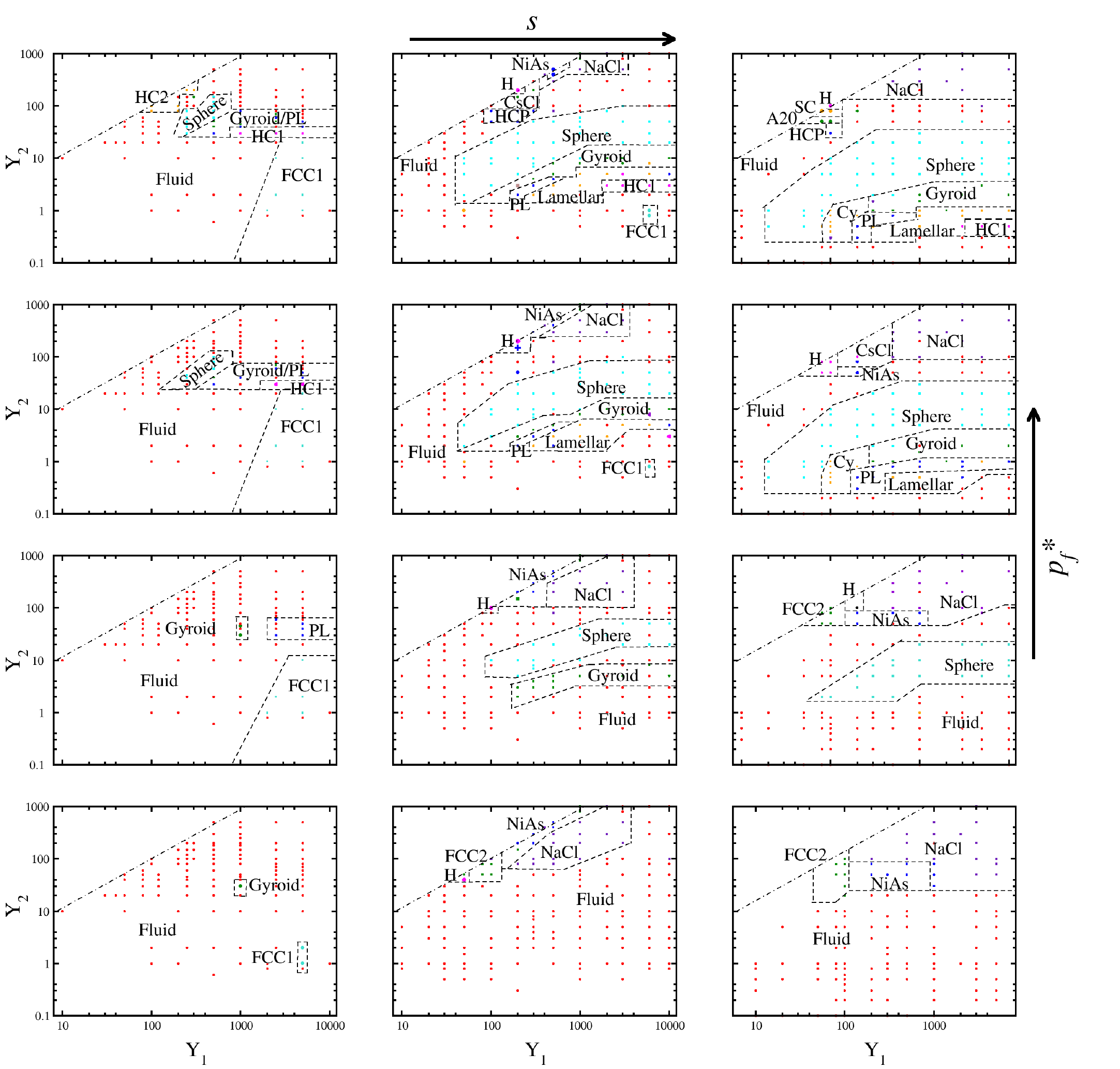}
\caption{$Y_1-Y_2$ phase diagrams of Hertzian dumbbells for size ratios $s=\{2,3,4\}$ and rescaled pressures $p_{f}^{*}=\{170, 510,1020,1700\}$. 
Crystalline phases are obtained for  $Y_1 \sim Y_2\gg1$, and include  NaCl-type, NiAs-type and crystals of the large dumbbellar component, FCC2, which melt and recrystallize first into H and CsCl-type crystals and then into SC, A20 and HCP structures as the pressure is increased.
 For $Y_1\gg Y_2\sim 1$ crystals of the small component, FCC1, are also observed. The region in between is dominated by mesophases including: FCC ordered spherical micelles (Sphere), hexagonally packed cylinders (Cy), gyroid and double gyroid (Gyroid), lamellar (Lamellar) and perforated lamellar (PL), honeycomb of the small components and cylinders of the large ones (HC1) and its inverted phase (HC2). }\label{hertz}
\end{figure}

The pressure-structure dependence of these crystals is reported explicitly in Fig. \ref{hertz} for $s=2, 3$ and $4$, and can be found near the upper boundary of the diagrams where $Y_1\simeq Y_2$.
Crystal structures were identified by combination of visual inspection, spherical harmonic-base order parameter $q_6$ ~\cite{ronchetti}, and by comparison with  the known crystal structures reported in~\cite{hudson, dijkstra1, dijkstra2,murray,web_lattice}. 
The most abundant one is the NaCl-type, where the large spheres form an FCC lattice and the small ones fill the octahedral holes, as an interpenetrating FCC lattice. Next  is the NiAs-type, where the large spheres pack in an HCP lattice and the small ones again fill the octahedral holes, this time in form of an interpenetrating simple hexagonal lattice. Coexistence of NaCl/NiAs-type crystalline phases in the AB-type mixtures has been observed experimentally and theoretically also in~\cite{dijkstra1,hunt}. The third crystal type is the FCC lattice of the large spheres, but no order for the small ones. The latter only develops for $s\geq3$ and values of $Y$ that are sufficiently small to allow hopping of the smaller components between different octahedral holes.
Further reduction of the Young's modulus leads to the destabilization of any crystal structure as the system approaches the ideal gas limit.

Figure \ref{hertz} shows twelve two  dimensional slices of the whole diagram, for a combinations of three size ratios and four different pressures. Depending on the combination of the four parameters we find a plethora of different phases, making the asymmetric case in which
($Y_2\ll Y_1$) the more interesting one. In fact, a whole new class of phases becomes accessible to the asymmetric dumbbells, including spherical and cylindrical micelles, gyroid, lamellar,  perforated lamellar and honeycomb phases. 
These phases are regularly observed in amphiphilic and diblock copolymer systems \cite{colloid_book, bates}, as well as in the polymer-nanoparticle tethered systems
\cite{amphi_book,glotzer1,glotzer2}. However, the aggregation in those systems is driven by both attractive and repulsive forces, while in our system no attractive force is present.

The symmetry/nature of these mesophases can be understood by analogous arguments 
 developed to describe the packing of amphiphilic molecules, i.e. the shape or more precisely the effective shape
 of the dumbbells determines how they will pack in three dimensions. Indeed, shape asymmetry is key to
 the formation of mesophases. Shape asymmetry can be obtained in our system by either setting a different size for the two components  constituting the dumbbell or by allowing one component to be softer than the other.  
Mesophases are observed when both physical size and softness asymmetry are present. 
We believe that this is because hard asymmetric dumbbells impose  packing constraints that are too restrictive   
for them to arrange into curved surfaces without creating  large stresses across the overall aggregate. 
This causes hard asymmetric dumbbells to form either crystalline or disordered phases. However, once the 
larger component of the dumbbell becomes deformable, packing constraints are relaxed and the formation of curved surfaces, with minimal interfacial energy and maximal entropy,  becomes possible.
Specifically, we find that, as $Y_2$ is decreased, the small and harder spheres can readily assemble into spherical clusters with overall FCC symmetry or hexagonally-ordered cylindrical clusters; in both cases these clusters are stabilized and separated by a fluid layer formed
by the overlapping larger spheres.
  
As $Y_2$ is further decreased the large component becomes more deformable, effectively modulating the shape of the dumbbell, which 
becomes less conical. This leads to the formation of first a gyroid phase of the smaller components,  then a double gyroid phase (involving both components), and finally, as the effective shape of the dumbbell resembles more and more that of a  cylinder, we find a lamellar phase. The thickness and the
stacking of the lamellar phase depends on the specific values of $Y_1$, $Y_2$ and $p_{f}^{*}$.
  
If we now also allow the smaller component to be deformed, additional curvature and degrees of freedom are introduced into the system which can
now also form a perforated lamellar phase \cite{hamley}, obtained by interconnecting the soft layers through holes 
within the hard layers. 

Changing the size ratio of the dumbbell or modulating the relative strength of the Young's moduli of the two components is not the only route to 
phase change in these systems. As mentioned above, for Hertzian interactions, it is possible 
to induce the formation of different solid phases at constant $s$, $Y_1$ and $Y_2$ by increasing the system pressure. 
Apart from the obvious transition from fluid to solid at low densities, we find that pressure can also drive several order-to-order
transitions. For instance a gyroid phase can transform into a perforated-lamellar phase which can then turn into a 
lamellar phase and subsequently a honeycomb structure.
Again, for large values of $s$ the pressure does effectively 
alter the overall geometric shape of the dumbbell by allowing for a greater degree of overlap among the larger components, which results into predictable 
symmetry changes of the aggregates, similarly to what we have previously discussed.

The change of pressure introduces even more dramatic transformations in the $Y_1\simeq Y_2$ region of the phase diagrams. 
Here we observe multiple re-entrant melting and crystal-crystal transitions upon pressure increase. Low pressure NaCl-type, NiAs-type and soft-sphere-FCC crystals melt under higher pressure and renucleate as CsCl-type binary crystals, or simple cubic (SC) and simple hexagonal (H) monocrystals of the soft spheres. At even higher pressures these crystals melt and again renucleate as A20-type monocrystals of the large spheres, where the small ones remain fluid, or HCP crystals of the large spheres, where the small spheres slightly overlap and aggregate into cylinders that fill the octahedral holes. Due to the bounded nature of the Hertz potential, many other crystal types would surely be found at even higher pressure, but we chose to stop compressing the system at $p*=1700$.

To test the robustness of our results, we also explored how structure formation is affected by the specific shape of the potential.
Namely, we recompute the phase diagram for the case in which interparticle softness in described by a generalized inverse power law potential (GIPL).
In this case, the extent of particle overlap is tuned by changing the exponent of the power law 
\begin{equation}
U(r_{ij})=
\begin{cases}
k_{\rm  B}T \left [ \left(\frac{\sigma_{ij}}{r_{ij}}\right)^{n_{ij}}-1\right ], & r_{ij}\leq \sigma_{ij}\cr
0, & r_{ij}>\sigma_{ij} \cr
\end{cases}
\label{eq:IPL}
\end{equation} 
where $\sigma_{ij}= (\sigma_i+\sigma_j)/2$, and the exponent  $n_{ij}$ is defined as $n_{ij}=\frac{n_{i}+n_{j}}{2}$, which
simplifies to  the more conventional inverse power law form when $n_i=n_j$ and $\sigma_i=\sigma_j$.
In this case the hard-core limit is  recovered for $\left(n_{ij} \to \infty\right)$, and can be used to operationally define the diameter 
of the components to be equal to $\sigma_i$. Upon decreasing $n_{ij}$ we can continuously allow for ever larger degrees of softness. 
Notice that, unlike the Hertz potential, the GIPL potential always presents an infinite barrier against complete particles overlap.

For consistency with our previous results we restricted the computation of the phase diagram to $\sigma_1<\sigma_2$ and $n_1\geq n_2$. 
Specifically, we considered size ratios  2, 3 and  4, and for each $s$ we explored several values of the exponents with $n_1\in [1,50]$ 
and $n_2\in [1,7]$. This combination of exponents was selected in an attempt to find a set  of parameters that are compatible to those
of the Hertzian system in its most interesting region of the phase diagram, i.e. we explicitly avoid the hard core limit for both components.
All of our simulations are carried out using the same number of dumbbells at room temperature, and the system is equilibrated
using the  pressure-annealing scheme described earlier.

\begin{figure}[h!]
\center
\includegraphics[width=180mm]{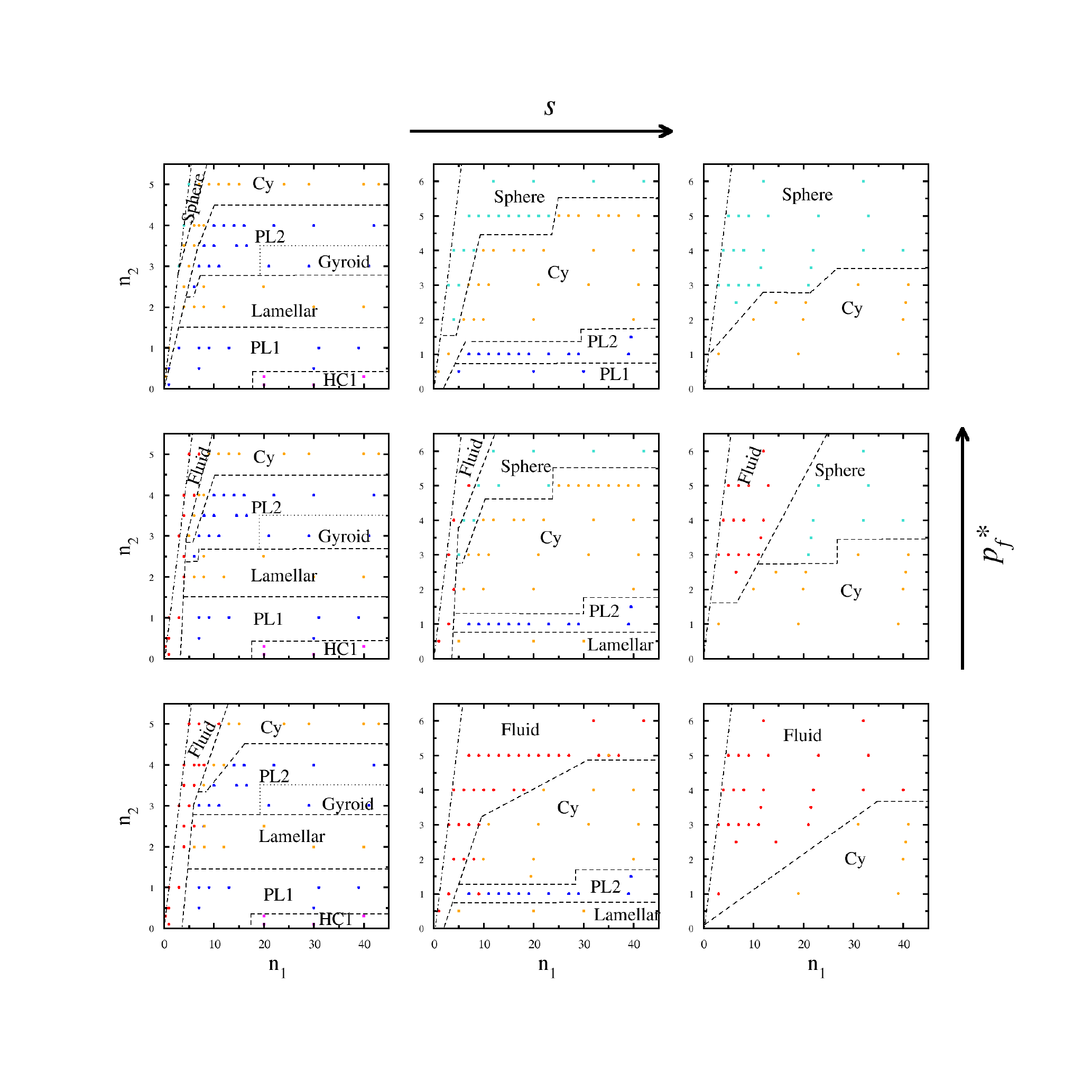}
\caption{$n_1-n_2$ phase diagrams of soft asymmetric dumbbells with GIPL potentials, for size ratios $s=\{2,3,4\}$ and rescaled pressures $p_{f}^{*}=\{86, 136,1200\}$. Qualitatively, the same mesophases as in the case of the Hertz potential are obtained. The perforated lameral phase appears here as two distinguishable phases $-$ perforated lamella through the layers of the large spheres (PL1) and perforated lamella through the layers of the small spheres (PL2).}\label{inv1}
\end{figure}

Figure \ref{inv1} shows the phase diagram of soft asymmetric dumbbells with GIPL potentials at three selected pressures and for the  three different size ratios. Here is a general summary of what we observe: At low pressures all systems are in the fluid phase. Upon increasing the pressure, each system phase-separates into ordered aggregates.  Similarly to what we observed for the Hertzian potential,  the form of these aggregates depends both on size ratio and relative softness of two components. Our results reveal the existence of several ordered phases  such as hexagonal cylindrical micelles, spherical micelles ordered into an FCC lattice, layered lamellar phases, perforated lamellar phase and a bicontinuous gyroid phase. Two types of the perforated lamellar phase are clearly distinguishable in different regions of the phase-space - those that with perforation  through layers of the small spheres and those with perforation through layers of the large spheres. In Fig. \ref{inv1} they are shown as separated phases. At intermediate pressures  despite the overall system symmetry, both components can freely diffuse within their own phase, while as pressure is increased crystallization of both components takes place.
 
\begin{figure}[h!]
\center
\includegraphics[width=100mm]{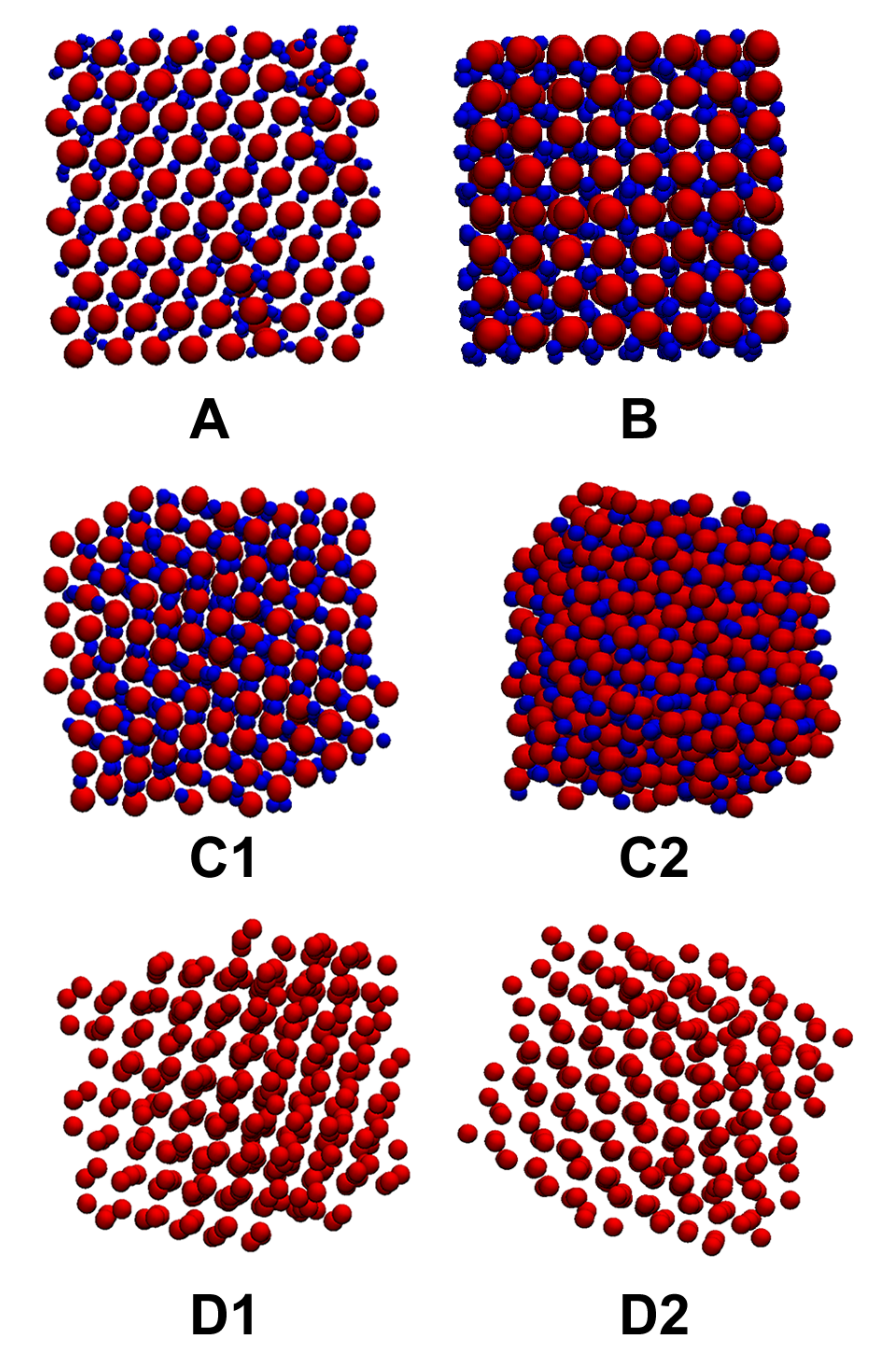}
\caption{Some of the crystals obtained in the system of the asymmetric dumbbells. The large sphere is depicted in red (light color) and the small one is depicted in blue (dark color):
(A) NaCl-type, (B) CsCl-type,\textbf{ (C1) and (C2) NiAs-type shown from two different perspectives, (D1) and (D2) A20 crystal shown from two different perspectives}. For the sake of clarity the sizes of the spheres has been altered. In snapshots (D1) and (D2) the small spheres remain fluid and are not shown.} \label{snapshots1}
\end{figure}

\begin{figure}[h!]
\center
\includegraphics[width=150mm]{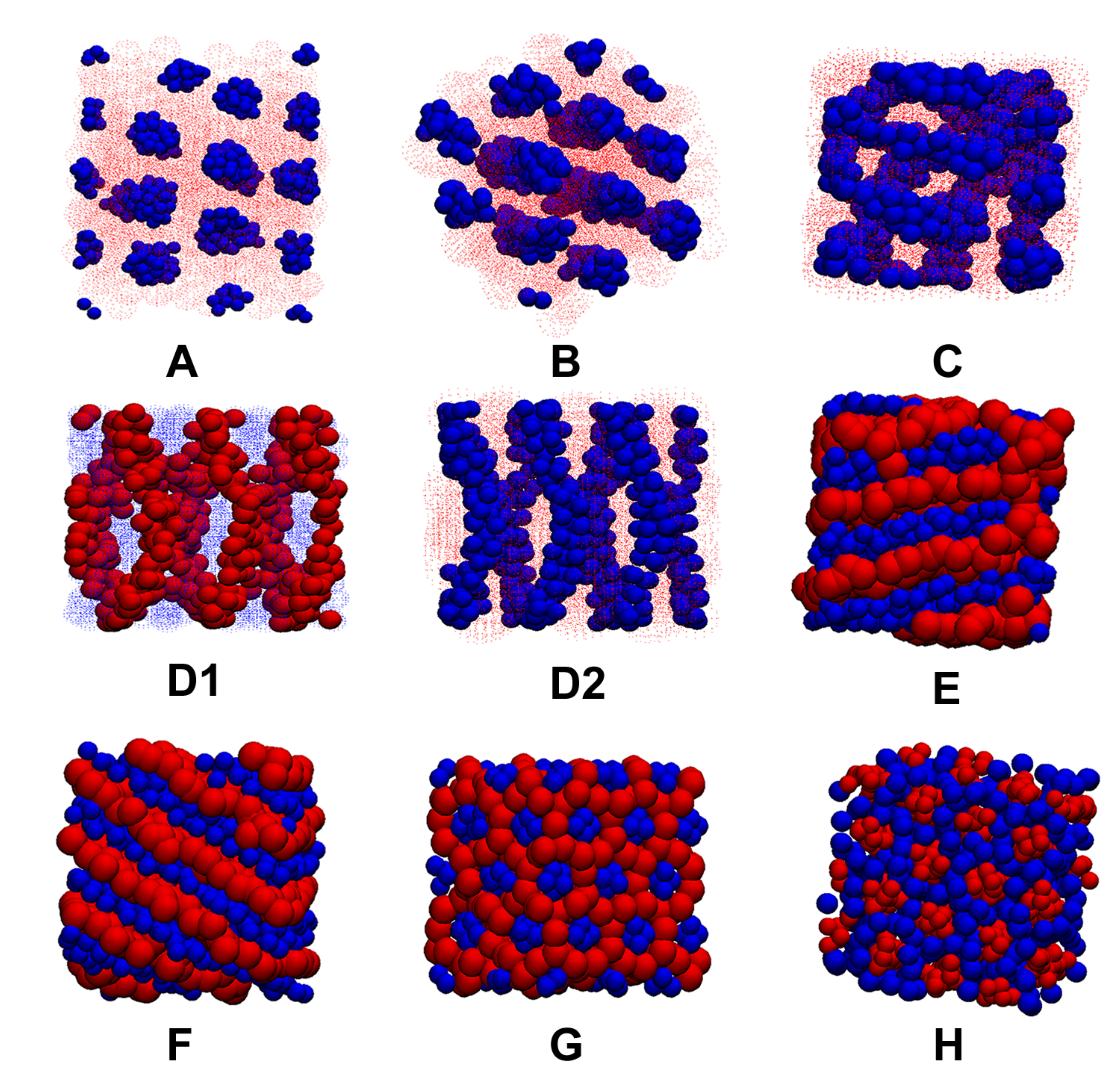}
\caption{Mesophases formed by a system of asymmetric dumbbells. Depicted in red (light color) is the larger and  softer component of the dumbbell,
while the smaller and harder  one is depicted in blue (dark color).  
(A) FCC ordered spherical micelle -Sphere (B) hexagonally packed cylinders -HPC, (C) gyroid of the hard spheres, (D1) and (D2) double gyroid phase -all referred to as Gyroid (E) perforated lamellar -PL, (F)  lamellar, (G) honeycomb of the soft spheres and cylinders of the hard ones -HC1, (H) honeycomb of the hard spheres and cylinders of the soft ones -HC2. For the sake of clarity the larger spheres are portrayed with a smaller diameter. In addition, in snapshots (A), (B), (C) and (D2) the soft spheres are depicted using a light, low density pixel representation, while in snapshots (D1) the same was done for the hard spheres.} \label{snapshots2}
\end{figure}

Although the details of the phase diagram are different from those resulting from the Hertzian dumbbells, it should be noticed that
using the  GIPL potential we managed to reproduce almost all of the aggregates previously obtained.
Due to our choice of exponents for the GIPL potential, a fair  comparison between the two phase diagrams should only include  
relatively small values of $Y_2$ in Fig.~\ref{hertz}.
The only  exceptions are the crystal of the smaller components FCC1, 
and the nature of the gyroid phase (indicated between dotted lines in the GIPL diagram) which seems to be metastable. 
We checked this by running several simulations with non identical initial conditions but under the same set of thermodynamic parameters.
We find that the gyroid phase does not consistently form as it instead does for Hertzian dumbbells. 
The metastability of the gyroid phase was discussed in detail for systems of  Block Copolymers in~\cite{gyro}
and for polymer tethered nanospheres in~\cite{glotzer2}.
A more specific study of this particular issue is out of the scope of our paper.
Figure \ref{snapshots1} shows snapshots of some of the crystalline phases found in this work, while Fig. \ref{snapshots2} shows snapshots of the mesophases obtained.

It is important to highlight that our results indicate that structure formation in these systems, within the range of pressures considered here, is remarkably robust with respect to the 
specific choice of the potential, be that bounded or unbounded. Nevertheless, apart from the similarities in the phase structure, there is a fundamental difference between the two systems which is most evident when both components are soft. Namely,   below a size-ratio-dependent value of $Y_i$ Hertzian dumbbells
remain in the fluid phase no matter how large is the external pressure imposed on the system, while the fluid phase for the GIPL potential
seems to always transforms into ordered aggregates once a sufficiently large pressure is achieved.
For instance, we  looked at the phase behavior of the GIPL system for $s=2$ and $n_1=n_2$ down to a value of the
exponent of  0.3, and we indeed keep on finding formation of ordered aggregates. This is obviously not conclusive, as it is not clear whether there exists an onset value of the exponent $n \geq 0$ (which could in principle be also size ratio dependent) 
below which the system stops crystallizing at any external pressure. This 
important but subtle issue is currently under investigation and the results will be published elsewhere. 
 
In conclusion, we have shown how the interplay between shape and deformability can be exploited to 
generate a large number of ordered mesophases in purely repulsive systems. 
We thoroughly explored the phase space for two cases of bounded and unbounded potentials covering a large  
range of systems spanning from the hardcore to the ideal gas limit (for Hertzian spheres). 
Our results show that the formation of most of the phases can be rationalized in terms of the 
effective geometry  of the components and are not very sensitive to the specific choice of the 
potential. Some of the phases that we can easily obtain with soft nanoparticles, namely the gyroid, double gyroid and perforated lamellar phases, are also observed in block copolymers and have been very recently the subject of intense investigation~\cite{glotzer2, hamley, gyro}. Nevertheless, bounded potential also give 
us access to multiple crystalline phases with pressure dependent symmetry,   effectively providing for
an almost unlimited variety of structures. Given the vast size of the phase space covered in this 
paper, we have not attempted to accomplish the heroic task of computing the equilibrium phase diagram and the relative stability of each of the many structures observed, but we hope that our results will inspire more theoretical and experimental work in this direction.
  
\section*{Acknowledgments}
This work was supported by the National Science Foundation
under CAREER Grant No. DMR-0846426, and partly by Columbia University.

\newpage
\

\

\

\

\

\
 
\newpage

\end{document}